\documentclass[twocolumn, pra,showpacs,nofootinbib,superscriptaddress]{revtex4}
\usepackage{dcolumn,graphicx}
\usepackage{amsmath}
\usepackage{amsfonts}
\usepackage{amssymb}
\usepackage{longtable}
\usepackage{epsfig}

\begin{document}
\title{ Dispelling the curse of the  neutron skin  in atomic parity violation}
\author{ B. A. Brown}
\affiliation{Department of Physics and Astronomy,
and National Superconducting
Cyclotron Laboratory,
Michigan State University,
East Lansing, Michigan 48824-1321, USA}
\author{  A. Derevianko }
\affiliation {Department of Physics, University of Nevada, Reno,
Nevada 89557}
\affiliation{ School of Physics, University of New South Wales, Sydney 2052, Australia}
\author{V. V. Flambaum}
\affiliation{ School of Physics, University of New South Wales, Sydney 2052, Australia}
\date{\today}
\begin{abstract}
We perform calculations for the neutron skin of nuclei and its contribution to atomic parity non-conservation (PNC) in many isotopes of Cs, Ba, Sm, Dy, Yb, Tl, Pb, Bi, Fr, Ra. Three problems  are addressed: i) Neutron-skin induced errors to single-isotope PNC,
 ii) Possibility to measure neutron skin using atomic PNC, iii)  Neutron-skin induced errors for ratios of PNC effects in different isotopes. In the latter
 case the  correlations in the neutron skin values for different isotopes lead to cancelations of the errors; this makes the isotopic ratio method a competitive tool in a search for new physics beyond the standard model.
\end{abstract}

\maketitle

Atomic parity non-conservation (PNC)
provides powerful constraints on extensions to the standard model
of elementary particles in the low-energy electroweak sector.
In such
measurements one determines a parity-violating signal $E_{\rm
PNC}$, related to the quantity of interest, the weak charge, $Q_W$, as $E_{\rm PNC}= k_{\rm PNC}\,
Q_W$. The coefficient $k_{\rm PNC}$ comes from
atomic calculations. Considering challenges faced by such calculations, an alternative approach  was proposed
by~\citet{DzuFlaKhr86}. The  idea was to form a ratio
$\mathcal{R}$ of the PNC amplitudes for two isotopes of the same
element. Since the factor $k_{\rm PNC}$ remains the same, it cancels out in the ratio.
However, \citet{ForPanWil90} pointed out a conceptual limitation to this approach -- an enhanced sensitivity of
possible constraints on ``new physics'' to uncertainties in the
{\em neutron} distributions. This problem is usually
referred to as the problem of the neutron ``skin''.
Almost for two decades this problem has persisted.
Here we show that the neutron skins in different isotopes are correlated; this leads to a substantial cancelation in the neutron skin induced uncertainties
in the PNC ratios. The use of modern experimental data and nuclear calculations
makes the isotopic ratio method a competitive tool in search for new physics beyond the standard model.

The neutron skin $\Delta R_{np}$ is defined  as a difference between the root-mean-square radii $R_n$ and $R_p$  of neutron and proton distributions.
Even in interpreting the most accurate to date single-isotope measurement in Cs~\cite{WooBenCho97}, this was a point of concern,
as the induced uncertainty was comparable to experimental error bar for the PNC amplitude\cite{PolWel99,VreLalRin00}.
The question was addressed in Ref.~\cite{Der02},
where  empirical antiprotonic-atom data fit for the neutron skin was used~\cite{TrzJasLub01},
and the associated uncertainty in the ``skin'' contribution to $E_{PNC}$ was substantially reduced.
At the same time this question is yet to be settled
for on-going PNC experiments with unstable analogs of Cs: Fr~\cite{GomOroSpr06} and Ra$^{+}$ \cite{WanVerWil08}.
In this letter we use results of recent advances in our theoretical
and experimental understanding of
neutron skins to address the important questions that can be
answered from atomic parity violation experiments.

%
%
%
%
%
%
%
%
%
%
%

\begin{figure}[h]
    \begin{center}
    \psfig{file=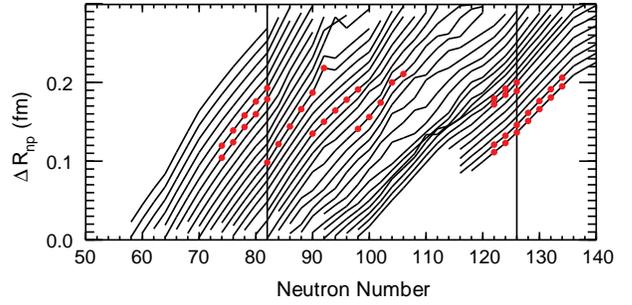,width=3.3in} 
    \end{center}
    \caption{The neutron skin $\Delta$R$_{np}$ for nuclei above
$^{100}$Sn are shown in the bottom panel (all $Z$, even $N$
values).
The nuclei for a given isotope are connected
by lines. The filled circles are those for the
nuclei of interest for atomic parity violation.
}
\label{s}
\end{figure}

{\em Nuclear-structure calculations--}
The nuclear structure models
used in the present calculations were constrained by
a number of measured observables, primarily by the
anti-protonic data. Until recently the anti-protonic atom data
were analyzed in terms of a simple Fermi shape
for the matter density. The anti-protonic data
are sensitive to the very low-density part of the
matter distribution, for $^{208}$Pb, for the density
near $R=10  $ fm \cite{BroSheHil07}. One cannot justify using the
simple
Fermi shape for the density distribution
at such large radii.

Recently the anti-protonic atom data have
been analyzed in terms of nuclear EDF models~\cite{BroSheHil07}
including those for the Skyrme and Relativistic Hartree
formulations. At large radii the matter density
is dominated by the neutron density (due to the
Coulomb barrier for protons). But the
connection between the asymptotic density
and the rms radius within the energy-density
functional models depends
on other features of models such as the
nuclear matter incompressibility $  K  $. The charge
density obtained from electron scattering
is best reproduced with EDF models with
$  K \approx 200-230  $ MeV. This eliminates many
of the parameter sets which give higher values of $  K  $.
Within the Skyrme models $  K  $ is closely
controlled by the power of the density-dependent
potential, $\rho^{ \alpha }$, with $\alpha$=1 with $  K \approx 330  $
down to $\alpha$=1/6  with $  K \approx 200  $.
The anti-protonic atom data were
compared to results of three new Skyrme
forces \cite{BroSheHil07}  called Skxsxx with $\alpha$=1/6 and
values of xx=15, 20 and 25
representing the neutron skin of $^{208}$Pb
in units of 10$^{-2}$fm.
The result of the analysis was
$  \Delta R_{np}=0.20(\pm0.04)(\pm0.05)  $ fm, where $  \pm 0.04  $ fm
is experimental error from the anti-protonic
line width, and where $  \pm0.05  $ fm
is the theoretical error
suggested from the comparison of the theoretical
and experimental charge densities at large radii.
The neutron skin can also be
constrained by the properties of the pygmy dipole resonance
in neutron-rich nuclei \cite{Pie06}.  Data for
$^{132}$Sn suggest a value of $  \Delta R_{np}=0.24(4) fm $ \cite{KliPaaAdr07}.
This is consistent with the value of 0.27(5) fm obtained
for $^{132}$Sn from the Skxs20(5) interactions.

The neutron skins for nuclei above
$^{100}$Sn obtained with Skxs20 are shown in Fig. 1.
These are obtained in a spherical basis with nucleons
allowed to occupy the lowest energy orbitals - the
calculation does not have deformation or pairing.
The irregularities between the neutron
magic numbers 82 and
126 are due to the filling of the proton h$_{11/2}$ orbital
and the neutron i$_{13/2}$ orbital. However,
since the orbitals in the major shells (g$_{7/2}$, d$_{5/2}$,
d$_{3/2}$, s$_{1/2}$, h$_{11/2}$) for protons and
(h$_{9/2}$, f$_{7/2}$,
f$_{5/2}$, p$_{3/2}$, p$_{1/2}$, i$_{13/2}$) for neutrons
are closely spaced, deformation and pairing will
average out these irregularities. Thus, in the discussion
of the $\mathcal{F}$ values
and their errors we will use
average values obtained from a smoothed variation
in the neutron-skin with neutron number. The
values can be improved with calculations
that include pairing and deformation, but our
conclusions should not change.
The error bars for the neutron skin were obtained from
[$\Delta$R$_{np}$(Skxs25) - $\Delta$R$_{np}$(Skxs15)]/2.

%

\begin{table}[h]
\caption{ Computed neutron skins $\Delta R_{np}$ and proton distribution rms radii. The relative contributions of the neutron skin
to atomic PNC amplitudes are listed in the last column.
\label{Tab:isotopes}}
\begin{ruledtabular}
\begin{tabular}{ccccc}
\multicolumn{1}{c}{A} &
\multicolumn{1}{c}{ $\Delta R_{np}$, fm} &
\multicolumn{1}{c}{$R_p$, fm} &
\multicolumn{1}{c}{$\delta E^{n.s}_{PNC}/{E_{PNC}}$} \\
\hline
\multicolumn{4}{l}{Cs ($Z=55$)}\\
129& 0.120(32)& 4.705& -0.0018(5) \\
131& 0.139(34)& 4.714& -0.0020(5)\\
133& 0.158(37)& 4.723& -0.0023(5)\\
135& 0.176(40)& 4.732& -0.0026(6)\\
137& 0.193(42)& 4.742& -0.0028(6)\\
\multicolumn{4}{l}{Ba ($Z=56$)}\\
130& 0.104(29)&  4.723 & -0.0016(4)\\
132& 0.124(32)&  4.731 & -0.0019(5)\\
134& 0.143(34)&  4.740 & -0.0022(5)\\
136& 0.161(37)&  4.749 & -0.0024(6)\\
138& 0.179(40)&  4.759 & -0.0027(6)\\
\multicolumn{4}{l}{Sm ($Z=62$)}\\
144& 0.098(27)& 4.894 & -0.0018(5)\\
146& 0.122(32)& 4.893 & -0.0022(6)\\
148& 0.144(35)& 4.903 & -0.0026(6)\\
150& 0.166(39)& 4.913 & -0.0030(7)\\
152& 0.187(43)& 4.924 & -0.0033(8)\\
154& 0.219(48)& 4.934 & -0.0039(9)\\
\multicolumn{4}{l}{Dy ($Z=66$)}\\
156& 0.135(35)& 4.997 & -0.0027(7)\\
158& 0.150(37)& 5.018 & -0.0030(7)\\
160& 0.164(38)& 5.039 & -0.0032(8)\\
162& 0.178(40)& 5.061 & -0.0035(8)\\
164& 0.191(42)& 5.082 & -0.0037(8)\\
\multicolumn{4}{l}{Yb ($Z=70$)}\\
168& 0.141(35)& 5.143 & -0.0031(8)\\
170& 0.153(38)& 5.163 & -0.0033(8)\\
172& 0.174(40)& 5.171 & -0.0038(9)\\
174& 0.202(51)& 5.173 & -0.0044(11)\\
176& 0.215(67)& 5.193 & -0.0046(14)\\
\multicolumn{4}{l}{Tl ($Z=81$)}\\
203& 0.179(45)&  5.422 & -0.0049(12)\\
205& 0.192(48)&  5.434  & -0.0053(13)\\
\multicolumn{4}{l}{Pb ($Z=82$)}\\
204& 0.172(44)& 5.430& -0.0049(12)\\
206& 0.184(46)& 5.442& -0.0052(13)\\
208& 0.200(50)& 5.450& -0.0056(14)\\
\multicolumn{4}{l}{Bi  ($Z=83$)}\\
209& 0.189(48)& 5.468 & -0.0054(14)\\
\multicolumn{4}{l}{Fr ($Z=87$)}\\
209& 0.121(36)& 5.518 & -0.0038(11)\\
211& 0.132(38)& 5.529 & -0.0041(12)\\
213& 0.146(42)& 5.536 & -0.0046(13)\\
215& 0.161(44)& 5.546 & -0.0050(14)\\
217& 0.176(47)& 5.555 & -0.0055(15)\\
219& 0.191(50)& 5.565 & -0.0059(16)\\
221& 0.206(53)& 5.574 & -0.0064(16)\\
\multicolumn{4}{l}{Ra ($Z=88$)}\\
210& 0.111(34)&  5.535 &-0.0035(11)\\
212& 0.123(37)&  5.546 &-0.0039(12)\\
214& 0.136(40)&  5.553 &-0.0043(13)\\
216& 0.151(43)&  5.563 &-0.0048(14)\\
218& 0.166(46)&  5.572 &-0.0053(15)\\
220& 0.181(49)&  5.581 &-0.0057(16)\\
222& 0.195(52)&  5.591 &-0.0062(16)\\
\end{tabular}
\end{ruledtabular}
\end{table}

{\em Atomic PNC and neutron skin --}
The PNC observables depend on matrix elements of weak interaction.
As demonstrated in Ref. \cite{PolForWil92},  the
matrix elements  of $H_{\rm W}$ may be parameterized as
\begin{eqnarray}
\langle j |H_W| i \rangle =                                  
\frac{G_F}{2 \sqrt{2}} \ C_{ji} \ R_{p}^{2 \gamma-2} \bar{Q}_{W} \, ,
\label{HW}
\end{eqnarray}
where factor $C_{ji}$ depends on atomic wavefunctions,
$\gamma = \sqrt{1 - (\alpha Z)^2}$ and $\bar{Q}_{W}$  includes
the dependence on nuclear distributions,
\begin{eqnarray}
\bar{Q}_{W}=                                                    
-N \, q_n + Z \, q_p \ (1 - 4 \,{\rm sin}^2 \theta_W) + \Delta Q_{\rm new}.
\label{QW}
\end{eqnarray}
The term $\Delta Q_{\rm new}$  characterizes ``new
physics'' and $\theta_W$ is the Weinberg angle.
The quantities $q_n$ and $q_p$ depend on the neutron and proton distributions
convoluted with atomic wavefunctions:
$
q_n = 1+f_n\left( \frac{R_n}{R_p}\right) \, .
$
In the ``sharp edge'' model of nuclear density distribution,
\begin{eqnarray}
f_n\left( \frac{R_n}{R_p}\right) \approx
- \frac{3}{70} \left( \alpha Z \right)^2
     \left[1+5 \left( \frac{R_n}{R_p} \right)^{2} \right] \,.        
\label{Eq:qn}
\end{eqnarray}
The accuracy of the above formula is sufficient for the goals of the present work.

{\em Single-isotope measurements -- }
The relative correction to the PNC amplitude due to the
neutron skin reads~\cite{Der02},
\begin{equation}
\frac{\delta E_{PNC}^{n.s.}}{E_{PNC}}=-\frac{3}{7}\left(  \alpha Z\right)
^{2}\frac{\Delta R_{np}}{R_{p}} \, .
\label{Eq:dEPNC-SingleIsotope}
\end{equation}
The computed corrections for all the isotopes are listed in Table~\ref{Tab:isotopes}.
In particular, for  $^{133}\mathrm{Cs}$ the relative correction is $-0.0023(5)$,
it is consistent with the value of -0.0019(8) from Ref.~\cite{Der02},  which was based on the semi-empirical
fit of antiprotonic-atom data~\cite{TrzJasLub01} ($\Delta R_{np} = 0.13(4)$ fm).
As we progress to heavier elements, the correction grows as $Z^2$,
reaching 0.6\% for Fr and Ra$^+$. As an example, for  $^{213}\text{Fr}$, the correction reads
$-0.0063(16)$. The error bar implies that at the present level of knowing the
neutron skin, it contributes to the uncertainty in  the extraction of new
 physics from the Fr experiment  at the 0.1-0.2\% level.

{\em Detecting neutron skin in atomic PNC ---}
The question of determining neutron skin is of interest in its own right,
for example, for equation of state for neutron stars.
It is worth mentioning the proposed
PREX experiment at JLAB \cite{prex} on using a PNC asymmetry in elastic scattering
of electrons from $^{208}$Pb to measure $  R_{n}  $ to
a 1\% ($  \pm  $ 0.05 fm) accuracy.  There have been
renewed attempts to obtain $  R_{n}  $ from hadronic
scattering data \cite{KarAmoBro02,ClaKerHam03}.

Considering this interest,
we would like to see if the neutron skin can be extracted from atomic PNC measurements. From the preceding discussion,
it is clear that  for the single-isotope PNC  the uncertainty of  experiments
and atomic calculations should be smaller than  0.2\% (Cs, Ba$^+$) and
0.6\% (Fr, Ra$^+$).
This seems to be a realistic goal~\cite{DerPor07,GinFla04}.

This problem may be also addressed in the isotopic chain experiments.
Suppose the PNC amplitudes $E_{PNC}$ and $E_{PNC}^{\prime}$ are measured for two
isotopes of the same atom or ion with neutron numbers $N$ and $N^{\prime
}=N+\Delta N$, and the ratio is formed
\begin{equation}
\mathcal{R}=\frac{E_{PNC}}{E_{PNC}^{\prime}}=\frac{\bar{Q}_{W}}{\bar{Q}_{W}^{\prime}%
}\left(  \frac{R_{p}}{R_{p}^{\prime}}\right)  ^{2\gamma-2}.\label{Eq:R}%
\end{equation}
Here all quantities with primes are for the isotope with $N^{\prime}$
neutrons. Focusing on the contribution of the neutron skin%
\[
\mathcal{R}\approx\frac{N}{N^{\prime}}\left(  \frac{R_{p}}{R_{p}^{\prime}%
}\right)  ^{2\gamma-2}\times\left(  1+\left[  f_{n}\left(  \frac{R_{n}}{R_{p}%
}\right)  -f_{n}\left(  \frac{R_{n}^{\prime}}{R_{p}^{\prime}}\right)  \right]
\right)
\]
If we neglect the difference between $R_{n}$ and $R_{p}$, i.e., the neutron
skin, then $
\mathcal{R}\rightarrow\mathcal{R}_{0}\equiv N/N^{\prime}\left(
{R_{p}}/{R_{p}^{\prime}}\right)  ^{2\gamma-2}%
$.
Any deviation of $\mathcal{R}$ from $\mathcal{R}_{0}$ is a signature of the
neutron skin. The figure of merit is
\begin{equation}
\Delta\mathcal{R}_{n.s.}   =\left(  \mathcal{R}-\mathcal{R}_{0}\right)
/\mathcal{R}_{0}=f_{n}\left(  \frac{R_{n}}{R_{p}}\right)  -f_{n}\left(
\frac{R_{n}^{\prime}}{R_{p}^{\prime}}\right) \, ,
\end{equation}
where $f_{n}$ is given by Eq.(\ref{Eq:qn}). Or in terms of the neutron skin
$
\Delta\mathcal{R}_{n.s.}\approx\frac{3}{7}\left(  \alpha Z\right)  ^{2}
\frac{1}{R_{p}}
\left[
\Delta R_{np}^{\prime}-\Delta R_{np} \right]  .
$

There are two observations that can be made: (i) The isotopic-ratios
are sensitive to the differential change in the skin thickness, i.e., the neutron-skin effects tend
to cancel and (ii) the largest effect is attained for a pair of isotopes where
the skin thicknesses differ the most. This  condition is reached for a pair
comprised of the lightest (neutron-depleted) and the heaviest (neutron-rich)
isotope of the chain.

For Cs, Ba, and Dy $\Delta\mathcal{R}_{n.s} \approx 0.001$,
$\Delta\mathcal{R}_{n.s}(\mathrm{Yb,Sm})%
\approx 0.002$, and $\Delta\mathcal{R}_{n.s}(\mathrm{Fr,Ra}) \approx 0.003$. Fr and Ra are the
extreme cases, as the skin thicknesses for the lightest and the heaviest
isotopes differ by a factor of two.
%

{\em Isotopic ratios: neutron skin vs. ``new physics'' --}
Suppose we form the ratio
of  measured PNC amplitudes for two isotopes of the same atom or ion.
Can we constrain new physics beyond the Standard Model from analyzing
the ratios? The previous studies have answered this question negatively,
as the uncertainties in the neutron skin masked the new physics contributions. Below, in light of our nuclear calculations, we re-visit this question.
The discussion follows previous analysis~\cite{DerPor02}.
There are two new points: (i) we use more finely-tuned calculations of the
skin (Table~\ref{Tab:isotopes} in lieu of the empirical fit of antiprotonic-atom
data~\cite{TrzJasLub01}) and (ii) we take into account that the errors in $\Delta R_{np}$
for two isotopes are correlated. Both these factors allow us to
argue that, by contrast to the previous studies, the isotopic ratios can provide competitive constraints on the new physics.

\begin{figure}[h]
    \begin{center}
    \psfig{file=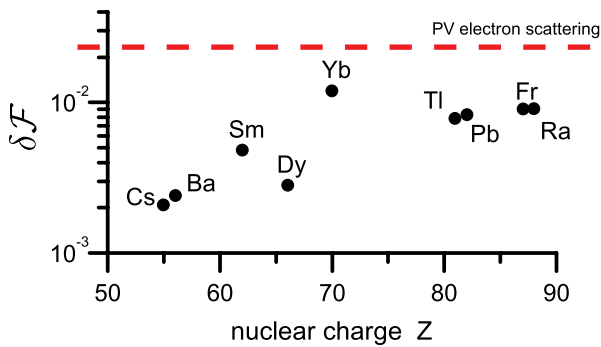,scale=0.7} 
    \end{center}
    \caption{Neutron skin vs. ``new physics''. The neutron-skin-induced uncertainties for
    isotopic chains are
    compared with the constraints from parity-violating electron scattering. }
\label{Fig:constraints}
\end{figure}

The term $\Delta Q_{\rm new}$ in Eq.~(\ref{QW}) characterizes ``new
physics'' at the tree level. Following~\cite{Ram99},
 we represent it as a combination of new-physics couplings to
protons and neutrons,$\Delta Q_{\rm new}  \equiv  Z \ h_p + N \ h_n $.
Various elementary-particle scenarios for
these interactions were reviewed in Ref.~\cite{Ram99}.
Then $\bar{Q}_W = N \ h_0 + Z \ h_p + N \ h_n$,
where $h_0$ comes from the Standard Model (this also includes the nuclear corrections). Unlike in the single-isotope measurements (sensitive mainly to $h_n$) in the isotopic ratio method, the sensitivity to new physics
comes predominantly due to $h_p$. The sensitivity may be parameterized
as~\cite{DerPor02}
\begin{eqnarray}
\mathcal{F}=\frac{ h_p}{h_0}= \left(  \frac{\mathcal{R}}{\mathcal{R}_0}-1 \right)
\frac{N\, N^\prime}{Z\, \Delta N} \, .                                  
\label{F}
\end{eqnarray}
In the absence of new couplings $\mathcal{F}=0$. The smaller $\mathcal{F}$,
the tighter the constraints on $h_p$ are. For a given chain, it is beneficial
to work with the largest possible neutron spread $\Delta N$, i.e.,
forming the pairs from the lightest and the heaviest elements of the chain.

The constraints on $h_p$, Eq.~(\ref{F}), are affected by (i) the experimental errors,
$\delta \mathcal{R}_{exp}$ and (ii) uncertainties in $\mathcal{R}_0$ which are
induced by insufficient knowledge of nuclear distributions.
Explicitly,
\begin{eqnarray}
\delta \mathcal{F}= \frac{N\, N^\prime}{Z\, \Delta N}
\left\{\frac{\delta \mathcal{R}_{exp}}{\mathcal{R}_0}
       + \delta (f_n - f'_n) \right\}.                            
\label{delF}
\end{eqnarray}
At this point we focus on the error in the  difference
\begin{eqnarray}
\delta (f_n - f'_n)
\approx \frac{3}{14}\left( \alpha Z\right) ^{2}\delta
 \left[ \frac{R_{n}^{\prime 2 }}{R_{p}^{\prime 2}}-\frac{R_{n}^2}{R_{p}^{2}}
\right]  \, .
\label{delf}
\end{eqnarray}
An important point is that the neutron-skin errors in Eq.(\ref{delf}) for two isotopes are correlated.
 From numerical experimentation, we find that for a given isotopic chain a variation of nuclear interaction parameters induces similar
change in the ratio $R_n/R_p$ for all the isotopes.
Indeed, in Table~\ref{Tab:isotopes} the boundaries of error bars in the values
of neutron skin correspond to the same values of nuclear parameters.
We see that the errors tend to cancel each other. This is to be contrasted
with the previous analysis of Ref.~\cite{DerPor02}: there the
errors from individual isotopes were treated independently, i.e.,
errors were added in quadratures. We find that our new ``correlated''
treatment reduces the error bars by four to ten times.

 We calculated $\delta
\mathcal{F}$  from a smoothed variation
in the neutron-skin with neutron number over pairs of the adjacent isotopes ($\Delta N$=2).
The results are compiled in Table~\ref{Tab:sensitivity}
and also in Fig.~\ref{Fig:constraints}. These errors are compared to
the new-physics couplings to protons, $h_p$, which are directly probed by the parity-violating
electron scattering (PVES). The PVES  experiments were recently analyzed in~\cite{YouCarTho07}.
The resulting  weak charge of the proton is $Q_{W}^p= 0.058 \pm 0.023$~\cite{YoungPrivate}. The error bar determines the upper bound on $\mathcal{F}$ shown in Fig.~\ref{Fig:constraints}.
It is clear that
all isotopic-chain determinations are competitive to bounds derived from PVES. For example,
measurements
with isotopes of Cs, Ba and Dy would be an order of magnitude  more sensitive
to the
new  physics.

\begin{table}
\caption{Contribution of nuclear-structure uncertainty to a constraint
on ``new physics'' $\delta \mathcal{F}$
 obtained from a smoothed variation
in the neutron-skin with neutron number. }
\label{Tab:sensitivity}
\begin{ruledtabular}
\begin{tabular}{cccc}
\multicolumn{1}{c}{Atom} & \multicolumn{2}{c}{Mass numbers $A$} &
\multicolumn{1}{c}{$\delta \mathcal{F}\times 10^3$} \\
\hline
Cs $(Z=55)$ & 129 & 137  &  2.1  \\
Ba $(Z=56)$ & 130 & 138  &  2.3  \\
Sm $(Z=62)$ & 144 & 154  &  4.2  \\
Dy $(Z=66)$ & 156 & 164  &  2.7  \\
Yb $(Z=70)$ & 168 & 176  &  10.2  \\
Tl $(Z=81)$ & 203 & 205  &  7.2   \\
Pb $(Z=82)$ & 204 & 208  &  7.7  \\
Fr $(Z=87)$ & 209 & 221  &  8.8  \\
Ra $(Z=88)$ & 210 & 222  &  8.9   \\
\end{tabular}
\end{ruledtabular}
\end{table}

AD was supported in part by the US Dept.~of State Fulbright fellowship to Australia. This work
 was supported in part by the National Science Foundation
grants No. PHY-06-53392 and PHY-0555366 and by the Australian Research Council.


\end{document}